\documentclass[aps,prb,reprint,groupedaddress]{revtex4-1}

\usepackage{graphics}

\begin{document}


\title{The 2D percolation transition at finite temperature:\\ The phase boundary for in-plane ferromagnetism in $\approx$ 2 ML Fe/W(110) films}


\author{R. Belanger}
\author{D. Venus}
\email{[corresponding author] venus@physics.mcmaster.ca}
\affiliation{Department of Physics and Astronomy, McMaster University, Hamilton, Ontario, Canada}

\date{\today}

\begin{abstract}
A  two dimensional (2D) percolation transition in Fe/W(110) ultrathin magnetic films occurs when islands in the second atomic layer percolate and resolve a frustrated magnetic state to produce long-range in-plane ferromagnetic order.  Novel measurements of the magnetic susceptibility $\chi(\theta)$ as the films are deposited at a constant temperature, allow the long-range percolation transition to be observed as a sharp peak consistent with a critical phase transition.  The measurements are used to trace the paramagnetic-to-ferromagnetic phase boundary between the $T=0$ percolation magnetic transition and the thermal Curie magnetic transition of the undiluted film.  A quantitative comparison to critical scaling theory is made by fitting the functional form of the phase boundary.  The fitted parameters are then used in theoretical expressions for $\chi(T)$ in the critical region of the paramagnetic state to provide an excellent, independent representation of the experimental measurements.
\end{abstract}


\maketitle

\section{Introduction}
Percolation phenomena play an important role in a diverse range of situations where connectivity across a macroscopic system is mediated by fragile, microscopic links.  As it is often the case that connectivity permits long-range order or non-equilibrium transport to exist, percolation is a central issue in condensed matter physics and materials science.   A few recent examples include the performance of field-effect transistors\cite{Aigner,Li}, the conductivity of silica powder\cite{Sokolowska}, and the metal-insulator transition\cite{Sherafati}.  There is, in addition, an enormous literature in computational simulations of percolation.

Classic experiments on 2D percolation at finite temperature were performed with neutron scattering on 3D crystals of quasi-2D diluted antiferromagnets\cite{Cowley,Birgeneau,Birgeneau2}.  Measurements of the correlation length of the staggered magnetization near percolation were used to study the multi-critical percolation point that can be approached either along the  concentration axis at $T=0$ or as a function of temperature very close to the percolation concentration.  These experiments, and subsequent theoretical work in critical scaling\cite{Coniglio}, established the important role of fluctuations of the fragile 1D links within clusters near percolation. The experiments demonstrated as well the difficulty in independently measuring the dilution concentration of the samples to sufficient accuracy to determine the critical exponents for percolation.  Of course, the use of crystals with fixed concentrations precluded the observation of the system as it actually went through the percolation transition.

An ultrathin film is a system where two-dimensional (2D) percolation is of obvious relevance to the structure of the film, as well as its static and transport properties.  Examples are homo- and hetero-epitaxial growth of films\cite{Chinta,Sattar} by the coalescence of islands, and the role of percolation in the transition from paramagnetism to ferromagnetism\cite{Elmers}.  A 2D magnetic film is in fact the prototypical system for the theoretical analysis of the 2D percolation transition and for the description of the phase boundary at finite temperature moving from the percolation point at zero temperature to a thermal Curie transition\cite{StaufferAharony}. 

 Experimentally, an ultrathin film is a true 2D system that is physically accessible, so that the fractional coverage $p$ of atoms on the substrate lattice can be changed and monitored in real time during the growth of the film.  It is therefore surprising that existing experimental studies of 2D percolation in magnetic films have not taken advantage of this accessibility, and have instead investigated a series of films of fixed composition in the vicinity of the percolation threshold\cite{Elmers,Bovensiepen,Kupper,Hope,Choi}.  These experimental investigations have used microscopy, magnetization measurements as a function of temperature, or both.  The studies are very successful at establishing whether the film is geometrically percolated or not.  In addition, the structural studies have provided important information about cluster distributions and their coalesensce, and the magnetic studies have plotted the line of Curie temperatures for percolated films as a function of the fraction of vacancies.  However, to our knowledge, quantitative real-time measurements of percolation itself have not been reported, nor has a quantitative comparison been made between critical scaling theory and experiment for the shape of phase boundary line between paramagnetism and (anti)ferromagnetism or the shape of the susceptibility curves themselves.

The current article presents magnetic susceptibility measurement of Fe/W(110) in real-time as the film is grown through percolation, and of samples of fixed deposition as a function of temperature.  The magnetic susceptibility is very sensitive to the phenomena of percolation, since it is a macroscopic property of the system that diverges in the paramagnetic phase near percolation (whereas the magnetization goes to zero).   Previous studies of Fe/W(110) films have shown that films grown at temperatures $T\geq600K$ exhibit step-flow growth\cite{Hauschild}.  At submonolayer depositions, the Fe forms disconnected monolayer ribbons, or strips, at the edges of monolayer terraces on the W(110) substrate.  Films prepared in this fashion have been used to study the crossover from 2D to 1D behaviour\cite{Elmers,Pratzer}.  If the films are grown instead at, or somewhat below, room temperature, the Fe forms islands, or clusters in the first atomic layer\cite{Elmers}.  As the deposition increases, the islands grow and begin to coalesce, and percolation of the first layer occurs when the first island spans the sample.  This behaviour has been observed experimentally through both imaging and the detection of long-range in-plane magnetization of the films\cite{Elmers3} near a deposition $\theta=0.60$ ML.  Further deposition creates 2nd layer islands of Fe on top of the first layer.  These islands have perpendicular anisotropy, perhaps because of magnetoelastic effects in the thicker strained, epitaxial films\cite{Weber}. This creates regions of fluctuating perpendicular magnetization within the in-plane magnetized first layer, and in this way destroys or frustrates long range order\cite{Elmers2}.  As the Fe deposition increases, the second layer islands grow and coalesce, and  long-range in-plane ferromagnetic order is  re-established abruptly upon percolation of the 2nd atomic layer.  A related jump in the critical temperature for ferromagnetism is seen in CoFe/C(001) films\cite{Bovensiepen} upon percolation of the second atomic layer.

The percolation of the 2nd atomic layer of Fe/W(110) is studied in detail in the present article.  Although we do not employ microscopy, we confirm a percolation transition that is robustly detected by magnetic susceptibility as the films are being grown.  The paramagnetic-to-ferromagnetic phase transition line measured in this way corresponds in quantitative detail to that predicted by critical scaling theories for percolation of a 2D Ising system.   Furthermore, the paramagnetic susceptibility measured as function of temperature near the phase transition line is described quantitatively by scaling theories that treat the percolation point at $T=0$ as a multicritical point.  

The remainder of this article is organized as follows. Section II briefly reviews the relevant critical theory for thermal and percolation transitions of an anisotropic Heisenberg model.  The next section describes the experimental methods for sample preparation and measurements of the magnetic susceptibility of the films in the present study.  The measurements are analysed in Section IV and fit to the critical scaling theory.  The conclusions are summarized in the final section.


\section{Theoretical background}
Fe/W(110) films near 2 ML in thickness grow with good layer completion\cite{Gradmann}, and have a strong in-plane magnetic anisotropy with an easy axis along the W[1$\overline{1}$0] direction.  This system is described by the 2D anisotropic Heisenberg model with nearest-neighbour exchange interaction $J$ and magnetic anisotropy $K$.  Without the anisotropy, the 2D Heisenberg model has no ferromagnetic ground state at finite temperature\cite{Herring}, but in the presence of the anisotropy, the system crosses over to a 2D Ising model\cite{Bander} as the correlation length $\xi$  diverges near $T_c$.  The bare 2D Ising model has transition temperature, $T_c^I$, where\cite{Kramers}
\begin{equation}
k_B T_c^I = 2.27 J.
\label{bare Ising}
\end{equation}
In the anisotropic Heisenberg model, spins wave fluctuations renormalize the exchange interaction to $J_{eff}(T)$.  According to first-order spin-wave theory\cite{Serena}, 
\begin{eqnarray}
J_{eff}(T) & = & J \biglb( 1- \frac{2T}{T_c^{sw}}\bigrb) \\
k_B T_c^{sw} & = & \frac{4\pi J}{\ln(J/K)}.
\label{sw}
\end{eqnarray}
Monte Carlo simulations\cite{Serena} have shown that a good estimate of the Curie temperature for the anisotropic Heisenberg model is given by substituting $J_{eff}(T=T_c)$ for $J$ in eq.(\ref{bare Ising}) and solving for $T_c$.  This results in
\begin{equation}
T_c \approx \frac{T_c^{sw} \: T_c^I}{T_c^{sw}+2T_c^I},
\label{T Heis}
\end{equation}
and yields, for example, $k_B T_c = 0.54 J$ when $K/J = 0.001$.  A number of experimental studies have confirmed that Fe/W(110) films near 2 ML thickness have a Curie temperature $T_c$ near 450 K, and critical exponents equal to the values of the 2D Ising model\cite{Back}; in particular, the critical exponent of the magnetic susceptibility was determined by direct measurement\cite{Dunlavy} to be, $\gamma = 1.75\pm0.02$.

The theoretical description of 2D percolation considers a 2D lattice with a fraction $0<p<1$ of the sites occupied by an atom.  As $p$ increases, the atoms are assigned randomly to unoccupied sites and clusters of maximal size $s$ are formed if $s$ occupied sites are connected by at least one path.  At the critical fraction, $p=p_c$, percolation occurs when at least one of the clusters becomes infinite in extent.  While this model does not describe how films usually grow, the two situations display the same universal behaviour in the critical region.  This is because in both systems the properties near percolation are determined by fragile 1D links that join 2D clusters into a single cluster.  As a result, there is a formal mapping of percolation on to a critical phase transition where properties scale as a power of $|p-p_c|$, and in particular to the paramagnetic-to-ferromagnetic transition of the Ising model with nearest-neighbour exchange\cite{StaufferAharony}.  In this correspondence, the magnetic correlation length, $\xi(p)$, is given by the correlation length for percolation, $\xi_p$, which is the average distance between sites within the same cluster.  The magnetic order parameter $M(p)$ corresponds to the ``strength" $S$, or fraction of the atoms contained within the infinite cluster.  The magnetic susceptibility $\chi(p)$ is determined by the average cluster size $\overline{s}$.  The universal behaviour of the magnetic system as it undergoes a percolation transition from paramagnetism to ferromagnetism can be expressed as
\begin{eqnarray}
\xi(p) & \sim & \xi_p \sim |p - p_c|^{-\nu_p} \\
\label{nup}
\chi(p) & \sim & \overline{s} \sim |p_c -p|^{-\gamma_p}.
\label{gammap}
\end{eqnarray}
Thus magnetic susceptibility measurements of a 2D film as a function of deposition (in principle at $T$=0), will exhibit a divergence due to percolation at $p=p_c$ with $\gamma_p = 43/18 \approx 2.39$, and measurements of a film without vacancies ($p$=1) as a function of temperature will exhibit a divergence due to a 2D Ising Curie transition at $T=T_c$ with $\gamma = 7/4$.

Scaling theory indicates that the percolation transition and Curie transition points are joined by a phase boundary for $p_c<p<1$.  The following description of the phase boundary between the paramagnetic and ferromagnetic states is adapted from ref.(\onlinecite{StaufferAharony}).  At low temperature near $p_c$, spin flips of the fragile 1D paths (or chains) that join together smaller clusters introduce a thermal correlation length $\xi_T$ that competes with $\xi_p$.  Rescaling the size of a 1D Ising chain  permits the identification of the thermal correlation length as\cite{Coniglio}
\begin{equation}
\xi_T\sim [\exp(\frac{-2J}{k_B T})]^{-\nu_T},
\label{xiT}
\end{equation}
where $\nu_T$ is the critical exponent for the thermal correlation length moving parallel to the temperature axis toward the percolation point at zero temperature.  Note that the exchange constant is the bare value, $J$, since it is due to localized 1D links that are not renormalized by the long-range spin wave excitations of the anisotropic Heisenberg model in the 2D clusters.  

The competition between the percolation and thermal correlation lengths modifies the critical susceptibility at finite temperature by a function $\chi_1[\frac{\xi_T}{\xi_p}]$, so that use of eq.(\ref{nup}) and (\ref{xiT}) yield
\begin{eqnarray}
\chi & \sim & |p_c-p|^{-\gamma_p}\:\chi_1(y) \\
y & = & \frac{\exp(\frac{-2J}{k_BT})}{|p_c-p|}.
\end{eqnarray}
This expression uses the result that   $\nu_p=\nu_T$ when clusters in the 2D Ising model are linked by 1D paths\cite{Coniglio}. The susceptibility will diverge along the phase boundary instead of at $p = p_c$ only if the function $\chi_1$ has a pole at some specific value $y=y_c$.  Then the equation of the phase boundary is given by the values $p^*(T)$ that satisfy
\begin{equation}
p^*(T)-p_c = \frac{1}{y_c}\exp(\frac{-2J}{k_BT}),
\label{boundary}
\end{equation}
where $p^*(T)\geq p_c$.  Close to the phase boundary, the standard ansatz gives $\chi_1\sim |y-y_c|^b$.  Because of the form of $y$,
\begin{equation}
y-y_c=\frac{y_c}{|p-p_c|}(p^*(T)-p_c+p_c-p).
\label{yc}
\end{equation}
For measurements at constant temperature,
\begin{equation}
\chi = \chi(p) \sim \frac{|p-p_c|^{-\gamma_p}}{|p-p_c|^b} |p^*(T)-p|^b,
\end{equation}
so that there is a transition at finite temperature $T$ on the phase boundary (and not at $p=p_c$) only if $b=-\gamma_p$.  For measurements in the paramagnetic region at  constant $p<p_c$, eq.(\ref{yc}) can be re-organized to give 
\begin{equation}
\chi = \chi(T)\sim |\frac{\exp(\frac{-2J}{k_BT})}{y_c}+(p_c-p)|^{-\gamma_p}.
\label{chipara}
\end{equation}
This temperature dependence reflects the fact that the phase boundary bends further from $p=p_c$ as the temperature increases.  Thus, at low temperature, one expects the critical exponent for percolation, $\gamma_p$, whether one approaches the phase boundary along a line of constant $p$ or constant $T$.  

Finally, as real films often do not grow as a single atomic layer, it is usually not possible to determine the fractional coverage $p$ of the relevant percolating layer as the film is being grown.  Rather, the experimental variable is the deposition, $\theta$, in ML.  $\theta$ is the fractional number of complete layers that the deposited atoms would form, \textit{if} each atomic layer was completed before the next began.  Given the universal properties of percolation, the small range in fractional coverage within the critical region, and the experimental ability to deposit films at a controlled, constant rate, the effect of substituting $\theta$ and $\theta_c$ for $p$ and $p_c$ in the above relations is simply to introduce a constant of proportionality that is absorbed in the prefactors to the power law scaling. 

\section{Experimental methods}

The Fe films were grown on a W(110) single-crystal substrate in an ultrahigh vacuum system.  The substrate was cleaned by oxygen treatments and flashing, and cleanliness was confirmed by Auger electron spectroscopey (AES) and low energy electron diffraction (LEED).   The substrate sample holder\cite{Venus} allowed either electron beam or radiative heating and cooling from a liquid nitrogen reservoir.  In this way, the temperature of the substrate, as measured using a W-Rh thermocouple embedded in the substrate crystal, could be maintained at a constant value or increased slowly at a constant rate.

The Fe was deposited by thermal evaporation from a pure wire, using electron bombardment\cite{Jones}.  The evaporated atoms passed through a pair of collimating apertures and impinged on the substrate.  The location and uniformity of the deposited film on the substrate were measured by AES, and adjusted using micrometers controlling an angular motion attaching the evaporator to the vacuum chamber.  The second collimating aperture on the evaporator was electrically isolated and the current due to Fe ions intercepted by it was measured using an electrometer.  This current, typically in the nA range, is a proportional measure of the Fe flux, and was kept constant during deposition using small adjustments of the position of the Fe wire.   The flux was calibrated using a series of sequential depositions on the W(110) substrate, annealing each deposit to 600 K to promote wetting, and measuring the attenuation of the AES signal from the tungsten substrate.  A plot of the AES signal against deposition time shows a clear change of slope at the completion of 1 ML, and allows the calibration of the flux current in nA min/ML\cite{Fritsch}.  The error in the calibration is 5\%.  A previous study has shown that the flux calibration is linear in ion current, and stable in time over the time periods required to grow films while the magnetic susceptibility is being measured in real time\cite{He}.  For this type of measurement, the uncertainty along the deposition axis derived linearly from the deposition time is entirely accounted for by the uncertainty in the original flux calibration using AES.

The magnetic susceptibility of the films was measured using the magneto-optic Kerr effect (MOKE)\cite{Arnold}.  A HeNe laser beam passed through a polarizing crystal, a UHV window, scattered from the sample at $45^o$, exited through a second UHV window, passed through a second polarizing crystal nominally crossed with the first, and was detected by a photodiode.  A pair of current coils produced an a.c. magnetic field at 210 Hz that was aligned with both the in-plane magnetic easy axis and the scattering plane of the laser.  The intensity changes in the photodiode signal, induced by optical rotation in the magnetic film \textit{via} the longitudinal Kerr effect, were detected using a lock-in detector.  This gave the susceptibility directly in nrad/Oe.  Optical compensation for the UHV windows provided sensitivity of 10 nrad/Oe for the susceptibility measurements.  The real and the imaginary (dissipative) parts of the susceptibility were measured simultaneously using the in- and out-of-phase components of the  signal from the lock-in amplifier.

Two types of measurements were made.  For the measurements of $\chi(T)$ at constant $\theta$, films were grown at room temperature and were not explicitly magnetized.  They were subsequently cooled to near 120 K and the susceptibility was measured in a small, 0.7 Oe a.c. field as the temperature was raised at 0.2 K/s to 400-500 K.  Previous studies have shown that a field of this amplitude produces a linear response in the critical region\cite{Dunlavy}.  Since the temperature during the measurement exceeded the growth temperature of the film, it is possible that some annealing of the sample occurred.  This is expected to be a minor effect, since in this system effective annealing resulting in important structural changes and wetting does not occur until at least 600 K.\cite{Elmers,Dunlavy}  For the measurements $\chi(\theta)$ as a function of deposition, the substrate was cooled to, and maintained at, the constant target temperature.  After the oscillating magnetic field was started, the shutter on the evaporator was opened and the susceptibility was recorded as a function of time.  The evaporator monitor current was maintained at 0.35 nA, which corresponds to about 5 min/ML.  Because of the short time that the films were exposed to the residual gas in the UHV chamber, the magnetic reorientation due to gas adsorption observed in previous studies\cite{Durkop} is not a factor in the current study.   Films were grown and simultaneously measured for a range of constant temperatures from 160 to 450 K.  Again, it is not expected that the film growth will be substantially different across this temperature range, but it is important to keep this possibility in mind when analysing the data set.

%


\section{Real-time measurements of percolation in a 2D magnetic film}

\begin{figure}
\scalebox{.6}{\includegraphics{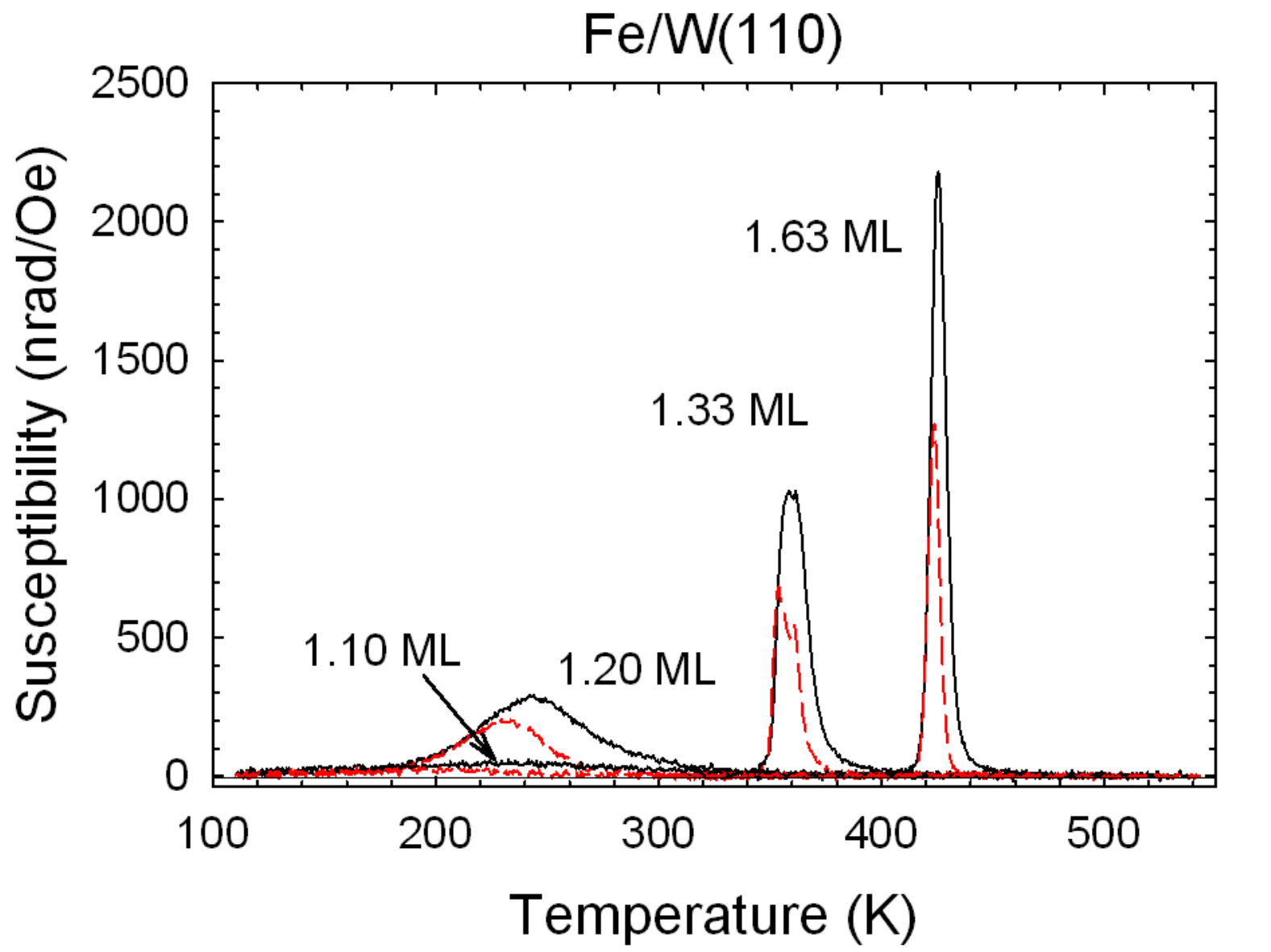}}
\caption{The magnetic susceptibility measured as a function of temperature for a series of films with different total deposition of Fe on W(110).  Each curve is labelled by the amount deposited.  The real part of the susceptibility is indicated by the solid line, and the imaginary part is indicated by the dashed line. }
\label{chiT}
\end{figure}

The study began with a series of measurements of $\chi(T)$ for films with fixed deposition, in order to confirm the presence of the percolation transition and determine its location in the phase diagram.  A selection of the measurements are shown along a common temperature axis in fig.(\ref{chiT}).  Measurements for films with $\theta<1.00$ ML Fe (not shown) have no response at the noise level;  a 1.00 ML Fe film has a very broad response just above the noise level that would not be visible on the scale of the figure.  The response peaks up significantly for the 1.20 ML film.  At 1.33 Fe ML the narrow peak in the susceptibility is qualitatively consistent with a critical phase transition.  The measurement  of the 1.63 ML Fe film has a narrow, well-defined peak with width $\Delta T_{FWHM}/T_c$ comparable to the width  in previous reports of the 2D Ising Curie transition for films near 2.00 ML\cite{Dunlavy,Fritsch}.  The maximum of the imaginary response is at a lower temperature than that of the real response, indicating a ferromagnetic to paramagnetic transition as temperature is increased.  These data suggest that a percolation transition may occur near to $\theta$=1.20 ML Fe. 

\begin{figure}
\scalebox{.5}{\includegraphics{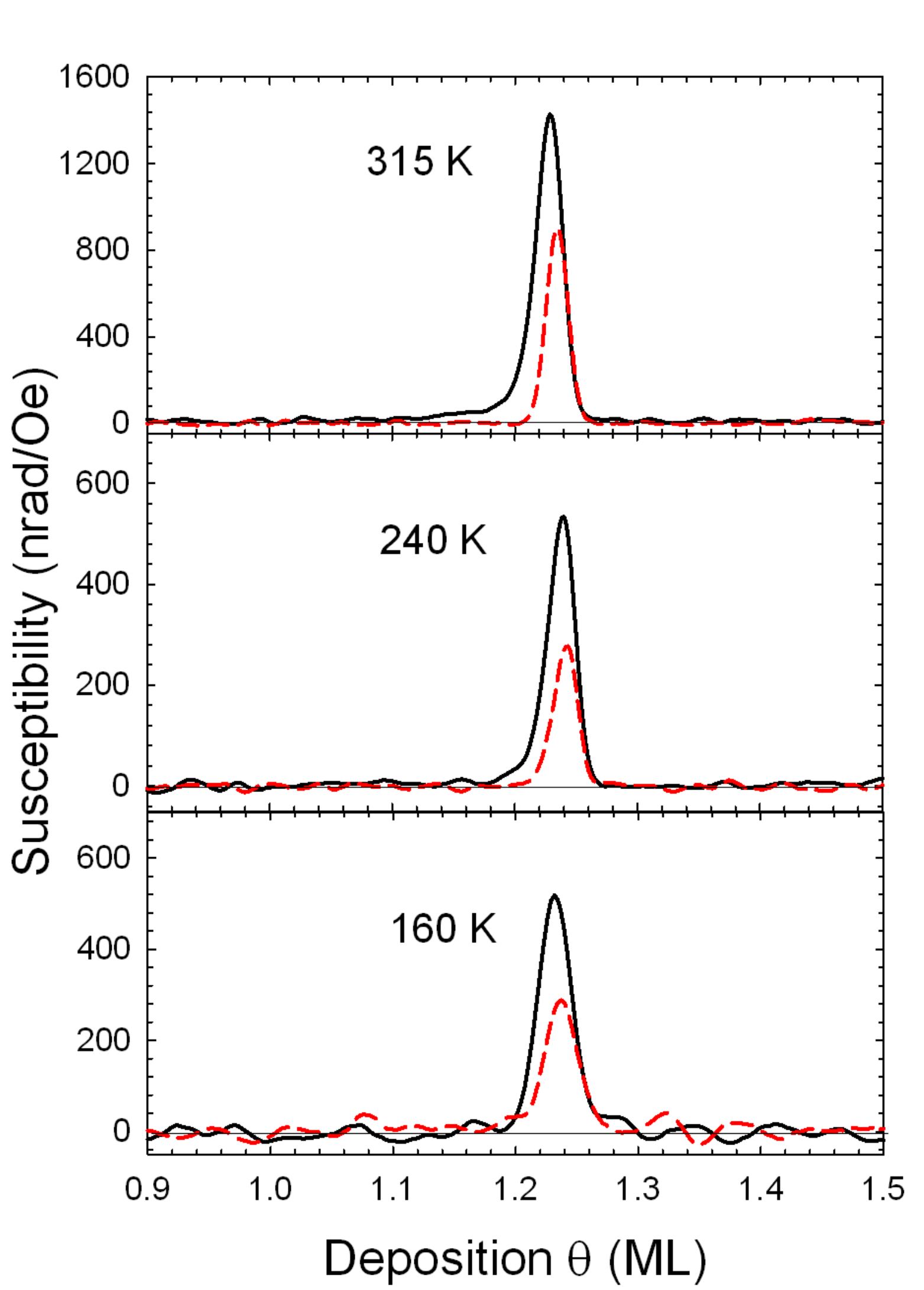}}
\caption{A representative sample of the magnetic susceptibility measured as a function of deposition while the Fe/W(110) films were being grown at constant temperature.  The real part of the susceptibility is indicated by the solid line, and the imaginary part is indicated by the dashed line. The data has been binned in increments of 0.004 ML.}
\label{chip}
\end{figure}

\begin{figure}
\scalebox{.8}{\includegraphics{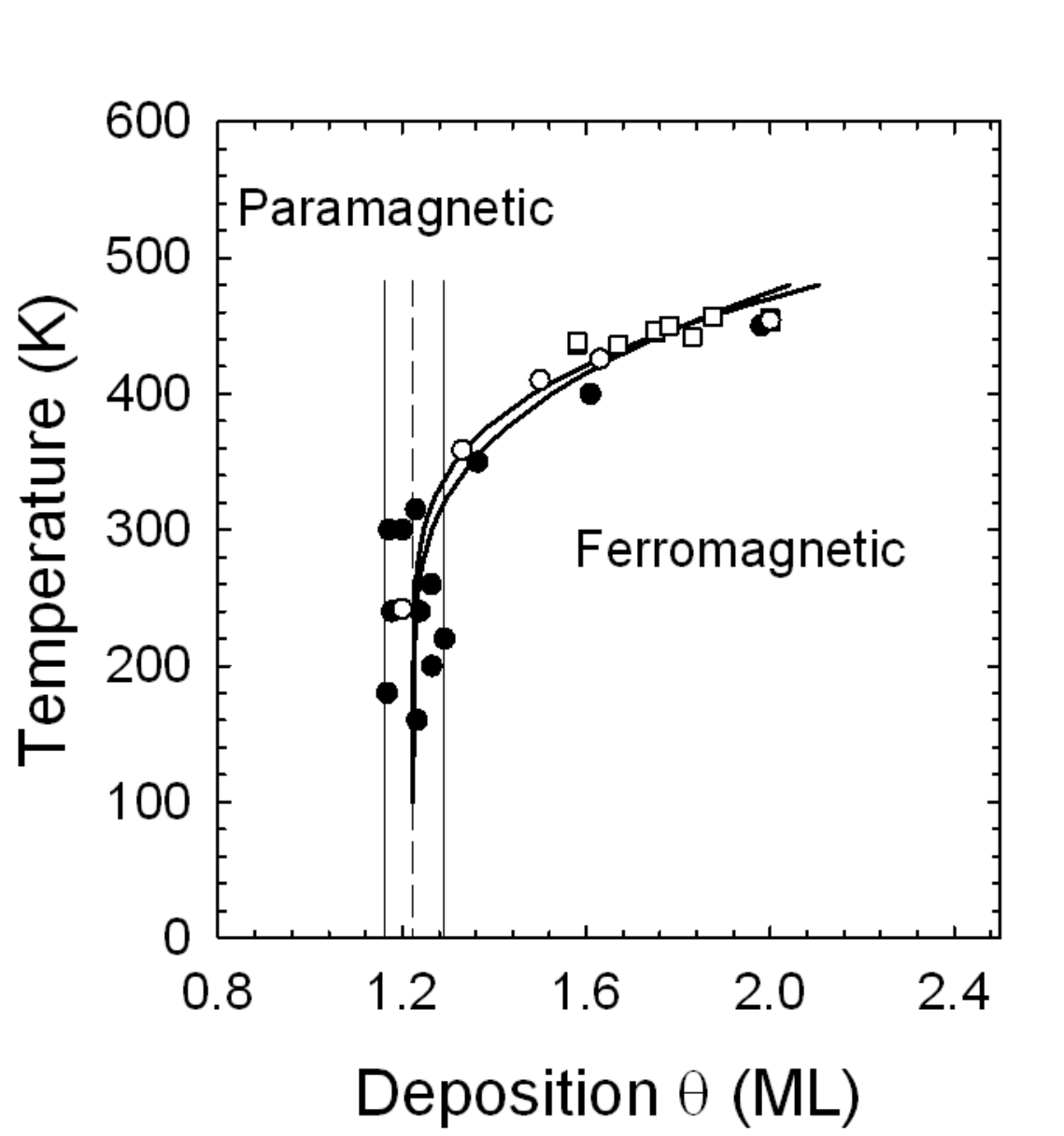}}
\caption{The temperature and deposition of the peak of the real part of the susceptibility is marked by symbols for a larger number of individual films.  The solid circles are for measurements at constant temperature as a function of deposition.  The open circles represent measurements of films with a constant deposition as a function of temperature.  The open squares are measurements as a function of temperature reported in ref.(\onlinecite{Fritsch}).  The phase boundary lines are fits to eq.(\ref{boundary}) for the limiting cases of the uncertainty bounds of the fitted parameters.}
\label{Tvsp}
\end{figure}

A representative sample of susceptibility traces measured at constant temperature as the films were being grown are shown in fig.(\ref{chip}).  The susceptibility has a remarkably narrow peak just above $\theta$=1.20 ML Fe, regardless of temperature in this temperature range (the susceptibility has no response above the noise level below 0.90 ML).  The peak in the imaginary response is at higher deposition than the peak in the real response, indicating a paramagnetic to ferromagnetic transition as the deposition is increased.  The results of measurements on a large number of films is summarized in fig.(\ref{Tvsp}).  Each symbol marks the temperature and deposition at which the real part of the susceptibility has a maximum for the individual samples.  Solid circles represent measurements made at constant temperature as the films were being grown, as in fig.(\ref{chip}).  Open circles represent measurements made on samples of fixed deposition as the temperature was changed, as in fig.(\ref{chiT}).  Open squares represent measurements made as a function of temperature that were reported in an earlier publication\cite{Fritsch}.  The peak positions for films grown at different temperatures and exposed to different temperatures during the measurement of the susceptibility are in good agreement, especially when the 5\% uncertainty in deposition due the AES calibration is considered.  This confirms that, in this system, magnetically relevant structural changes due not occur until temperatures above at least 460 K.

The phase diagram in fig.(\ref{Tvsp}) is qualitatively consistent with a magnetic transition due to percolation of islands in the 2nd atomic layer.  An essential point is that for measurements at constant temperature (closed circles) below 315 K, the deposition at the peak of Re $\chi(\theta)$ is independent of temperature within the uncertainty due to the deposition calibration.  The dashed vertical line indicates the average of these points, and is interpreted as $\theta_c=1.224\pm0.043$ ML.  The solid vertical lines that mark $\pm$5\% from this average value include all the data points, showing that all of the variation is consistent with the deposition calibration.  As in many magnetic studies, the peak in the susceptibility at the transition is the most precise marker of thickness.  This consistency also shows that despite the fact that Fe/W(110) is a complicated magnetic system, the film growth is reproducible and measurements made on the different films represent a self-consistent data set.  

The quantitative fitting of the phase boundary line in fig.(\ref{Tvsp}) is discussed later.  Qualitatively, it is  consistent with the findings of Elmers \textit{et al.}\cite{Elmers2}, but there are some differences.  First, the onset of in-plane magnetization in the previous study was at 1.48 ML, whereas it is 1.224 ML here.  Given that the deposition at which 2nd layer percolates is dependent upon the growth, island structure and fractional completion of the first layer before the second layer begins, it is not surprising that the sudden appearance of long-range in-plane order varies somewhat from laboratory to laboratory as the fine details of the protocols and experimental conditions for film preparation vary.  For example, the scanning tunnelling microscopy images of percolated samples showing a Curie transition are quite different in ref.(\onlinecite{Elmers}) and (\onlinecite{Back}).  

A more surprising difference is the lack of a response in the magnetic susceptibility at the appearance of long-range in-plane ferromagnetism in the first atomic layer below 230 K near $\theta=0.60$ ML, as reported by Elmers \textit{et al.}\cite{Elmers,Elmers2}  This may be due to the different magnetic histories of the samples in the two investigations.  In the study that detected long-range ferromagnetic order in the first atomic layer, the films were deposited at room temperature and then cooled to 115 K in the presence of repeated 200 Oe field pulses.  This locked the film in a saturated remanent state as the coercive field increased rapidly below 230 K, and the remanent state was detected by spin-polarized electron scattering.  By contrast, in the current study, the films were grown at a low temperature in the presence of an a.c. field of only 0.7 Oe.   Before first layer percolation occurred, the independent islands would likely be blocked, unresponsive to the small field used for the susceptibility measurements, and the net magnetization, averaged over the islands, would be zero.  In this case, the multi-domain state formed upon first layer percolation of the blocked islands would be difficult to detect.  By contrast, second layer Fe islands on a continuous first layer would not be magnetically isolated, even though the system is magnetically frustrated, thus permitting the observation of a coherent magnetic state upon second layer percolation\cite{Elmers2}.

\begin{figure}
\scalebox{.5}{\includegraphics{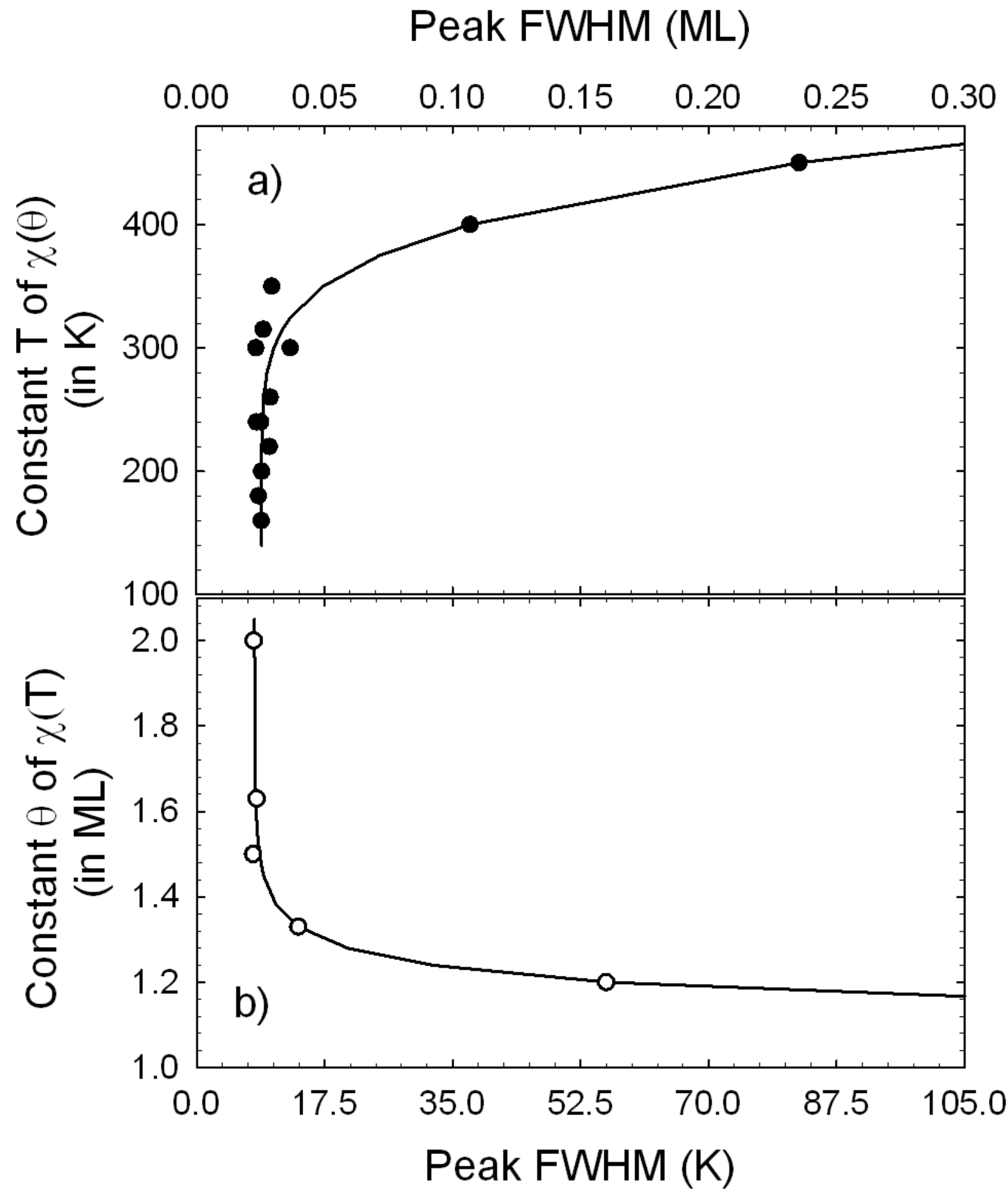}}
\caption{The full width at half maximum (FWHM) of the susceptibility peaks are plotted. a) The solid symbols plot the FWHM of $\chi(\theta)$ (top scale in ML) against the constant temperature at which the measurements were made as the Fe films were being grown.  b) The open symbols plot the FWHM of $\chi(T)$ (bottom scale in K) against the deposition of prepared films.}
\label{fwhm}
\end{figure}

Further qualitative evidence of the percolation transition is provided in fig.(\ref{fwhm}), where the full width at half maximum (FWHM) of the susceptibility peaks are plotted for the measurements made at constant temperature as the films were grown (part a)), and for the measurement made on films of fixed deposition as a function of temperature (part (b)).  The former illustrates that all of the measurements $\chi(\theta)$ as a function of deposition that have peaks at a consistent value of $\theta=\theta_c$, also have consistently narrow, sharp peak shapes consistent with a critical phase transition.  Constant temperature measurements at higher constant temperature are qualitatively different in that the FWHM increases.  Part b) of the figure illustrates a complementary situation for the measurements $\chi(T)$ for films with fixed deposition.  They have a consistent, narrow peak only for depositions above 1.5 ML.  The correspondence with fig.(\ref{Tvsp}) is obvious.  Crossing the phase boundary at high deposition as a function of temperature represents the limiting case of a Curie transition.  Crossing the phase boundary as a function of deposition at low temperature represents the limiting case of a percolation transition.

Alternate interpretations of the phase boundary are not consistent with the data.  The phase boundary determined by a cross-over from 2D films to 1D structures (such as isolated strips) has been previously determined experimentally for this system\cite{Elmers}.  It has a very different shape and extends down to $\theta$=0.05 ML.   A reorientation transition of second layer islands from perpendicular to in-plane magnetization without percolation might occur, driven by structural changes that relieve strain once the islands reach a critical size\cite{Elmers5}.  However, this type of transition would be very broad as a function of deposition, since the critical island size would sweep through a broad distribution of island sizes.  Although there are likely to be complex microscopic structural and magnetic origins of the frustrated magnetic state when $\theta<\theta_c$, it is clear that the frustration is removed by percolation of the second layer islands.  Because the percolation transition exhibits universal critical behaviour, this is the correct characterization of the transitions itself, as observed in $\chi(\theta)$ measurements at lower temperature.
\begin{figure}
\scalebox{.55}{\includegraphics{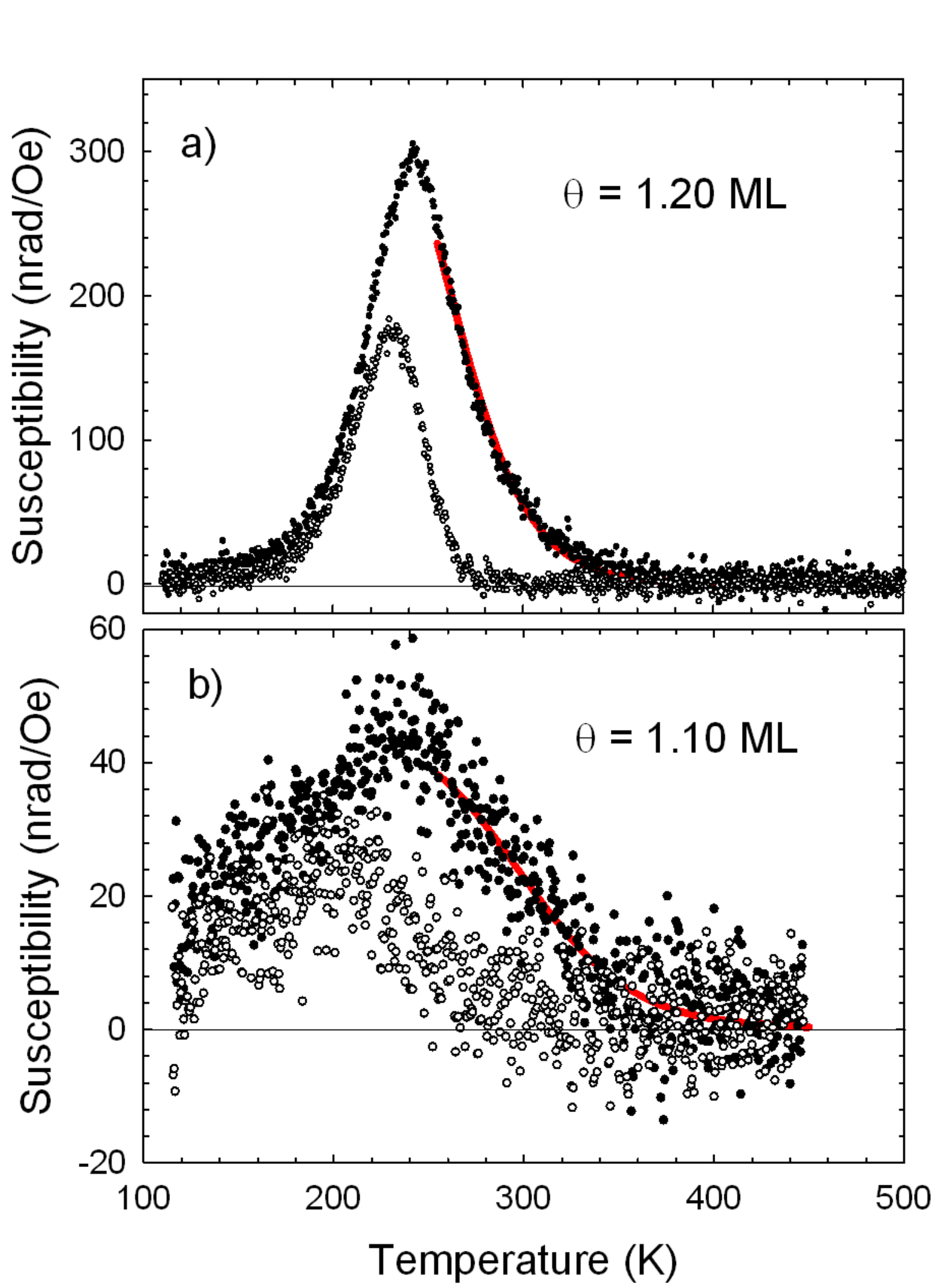}}
\caption{The measurements of $\chi(T)$ for films in the paramagnetic region close to the phase boundary are  reproduced on a larger scale.  Part a) shows data for $\theta$=1.20 ML, and part b) shows data for $\theta$=1.10 ML. The real parts are indicated by solid symbols, and the imaginary parts by the open symbols. The lines are fits of Re $\chi(T)$ to eq.(\ref{chipara}) at temperatures above 255 K, where the dissipation, represented by Im $\chi(T)$, is small.}
\label{1p2ML}
\end{figure}

The experimental results can be compared to the predictions of critical scaling theory for the phase boundary connecting a 2D Ising transition as a function of temperature and a 2D percolation transition as a function of deposition.   The data between the solid vertical lines in fig.(\ref{Tvsp})was used to determine $\theta_c$, and the remaining points were used in a least-squares fit to eq.(\ref{boundary}). This yields the parameters $\ln(y_c)=5.4\pm0.6$ and $2J/k_B=2670\pm270$ K. The two bold lines in fig.(\ref{Tvsp}) show the limiting cases for the phase boundary using the fitted parameters at their error bounds.  Using the measured value of the surface anisotropy\cite{Elmers4} $K=$ 0.61 mJ/m$^2$ gives $K/k_B\approx 1.4$ K/atom, so that $K/J\approx0.0005$ and the estimate just below eq.(\ref{T Heis}) is valid.  This gives an estimated transition temperature for the anisotropic Heisenberg model $T_c\approx670 K$, which is in  reasonable agreement with the experimental value 450 K for a 2 ML Fe/W(110) film.  

Revisiting fig.(\ref{chiT}) and fig.(\ref{Tvsp}), it can be seen that the measurements for the films with depositions $\theta$ = 1.10 and 1.20 ML represent the paramagnetic susceptibility measured close to the phase boundary, along a line parallel to the temperature axis that leads to the $T=0$ percolation point.  Eq.(\ref{chipara}) should therefore describe $\chi(T)$ so long as dissipative effects (as indicated by Im $\chi(T)$) are not important.  These data are reproduced on a larger scale in fig.(\ref{1p2ML}), where the solid symbols plot the real part of the susceptibility and the open symbols plot the imaginary part.  It can be seen that for these measurements made with an oscillating field frequency of 210 Hz, Im $\chi(T)$ sets a lower temperature limit of $T\approx255$ K for the region where the equilibrium response is measured.  

The parameters $J$ and $y_c$ required in eq.(\ref{chipara}) have already been determined from fitting the phase boundary in fig.(\ref{Tvsp}).  Beginning with the data for $\theta$= 1.20 ML, one expects $\theta_c-\theta\approx$ 0.024 ML, but it is difficult to estimate an error bound on this value.  The error bound is certainly less than the 5\% uncertainty in the deposition calibration, as this range would include negative values.  Also, it is not possible to determine $\gamma_p$ directly from the current set of measurements of $\chi(\theta)$.  Measurements at much lower deposition rates are needed to provide curves with more data points and a greater signal-to-noise ratio in the critical region. One way to proceed is to demonstrate that the data set is entirely consistent with the theoretical expectations.  If the theoretical value of $\gamma_p=43/18\approx2.39$ is assumed to be exact, a fit to eq.(\ref{chipara}) for the deposition yields $\theta_c-\theta=0.029\pm0.002$ ML.  The error bounds are determined by using the parameters from the two limiting phase boundaries in fig.(\ref{Tvsp}), since this is larger than the statistical error in each fit.  This is very close to the calibrated deposition, despite the large calibration uncertainty.  The line through the data in fig.(\ref{1p2ML}a) corresponds to this fitted function.

For the data with $\theta$=1.10 ML, $\theta_c-\theta\approx$ 0.124.  In this case, the combination 
$\exp(-2J/k_BT)/[y_c(\theta_c-\theta)] <<$1, and a linear expansion of eq.(\ref{chipara}) shows that the fitted curve depends only on $\gamma_p/(\theta_c-\theta)$, and not on each parameter individually.  If once again the theoretical value for $\gamma_p$ is assumed, then $\theta_c-\theta = 0.101\pm0.006$.  Again, the error bounds are determined by the parameters corresponding to the two limiting phase boundaries, and the fitted curve is shown on fig.(\ref{1p2ML}b).

The self-consistency of the entire data set and its agreement with the scaling theory for the 2D percolation critical transition of an anisotropic Heisenberg model is very encouraging.  Unfortunately, a complete experimental test is not possible without an experimental determination of $\gamma_p$.  One could assume that the deposition of $\theta$=1.20 ML in fig.(\ref{1p2ML}a) is exact and then fit for $\gamma_p=2.4\pm0.2$, but this seems to be of questionable value.  This same difficulty was encountered in the early studies of quasi-2D diluted antiferromagnets\cite{Cowley,Birgeneau}, where dilution concentrations more accurate than those found by chemical analysis of the samples were required to arrive at a value of $\gamma_p$ to compare to theory.  In that case, in order to make progress, the dilution concentrations were calculated using a method that assumed the theoretical critical exponent for the percolation correlation length, $\nu_p$, was exact.  However, in the present case, the accessibility of the true 2D film geometry and the demonstrated ability to measure the susceptibility as a function of deposition as the system percolates provides a path to accomplish this final step. The direct, experimental determination of the value of $\gamma_p$ will be the subject of a future publication.



%



\section{Conclusions}
Novel measurements of magnetic susceptibility as an ultrathin film is being grown have been used to study the 2D percolation transition in a true 2D film system, and have allowed a quantitative comparison to the predictions of scaling theory.  Fe/W(110) films exhibit a frustrated magnetic state when isolated 2nd atomic layer islands are formed on a more complete 1st atomic layer.  This is because the 2nd layer islands have perpendicular magnetic anisotropy, whereas the 1st layer has in-plane anisotropy.  The complicated frustrated state is removed at the percolation of the 2nd layer islands, and long-range in-plane magnetic order is established.

Measurements of $\chi(\theta)$ at constant temperature reveal a sharp peak at the paramagnetic-to-ferromagnetic transition caused by percolation.  The deposition, $\theta_c$, and peak width at percolation, is independent  of temperature over a substantial range of temperature, as is expected for a geometric  percolation  transition (as opposed to a thermal transition). Measurements of $\chi(T)$ at constant deposition confirm previous findings of a sharp peak that marks the thermal ferromagnetic-to-paramagnetic Curie transition.  

Scaling theory predicts the shape of the phase boundary between the percolation point at $T=0$ and the 2D Ising Curie transition of an undiluted film.  This prediction has been confirmed by plotting the transition peaks from a large set of both types of susceptibility measurements.  The experimental phase boundary conforms very well to the theory, and the fitted parameters have physically relevant values.  For example, the bare nearest neighbour exchange coupling is fitted as $J/k_B=1335\pm135$ K and gives an estimate of the Curie temperature of the anisotropic Heisenberg model of about 660 K, which is to be compared to the measured transition temperature of 450 K for 2 ML Fe/W(110).   The parameters from the phase boundary are used to subsequently fit the paramagnetic susceptibility $\chi(T)$ in the critical region using scaling theory for the 2D percolation transition of a 2D Ising system.  The only independent assumption that is required to provide excellent, self-consistent fits to the data is that the theoretical value of the percolation critical exponent, $\gamma_p$, is exact.  These experimental methods offer great promise for the future experimental determination of $\gamma_p$ and a completely experimental characterization of the 2D percolation transition.



\bibliography{percolation}

\begin{thebibliography}{36}%
\makeatletter
\providecommand \@ifxundefined [1]{%
 \@ifx{#1\undefined}
}%
\providecommand \@ifnum [1]{%
 \ifnum #1\expandafter \@firstoftwo
 \else \expandafter \@secondoftwo
 \fi
}%
\providecommand \@ifx [1]{%
 \ifx #1\expandafter \@firstoftwo
 \else \expandafter \@secondoftwo
 \fi
}%
\providecommand \natexlab [1]{#1}%
\providecommand \enquote  [1]{``#1''}%
\providecommand \bibnamefont  [1]{#1}%
\providecommand \bibfnamefont [1]{#1}%
\providecommand \citenamefont [1]{#1}%
\providecommand \href@noop [0]{\@secondoftwo}%
\providecommand \href [0]{\begingroup \@sanitize@url \@href}%
\providecommand \@href[1]{\@@startlink{#1}\@@href}%
\providecommand \@@href[1]{\endgroup#1\@@endlink}%
\providecommand \@sanitize@url [0]{\catcode `\\12\catcode `\$12\catcode
  `\&12\catcode `\#12\catcode `\^12\catcode `\_12\catcode `\%12\relax}%
\providecommand \@@startlink[1]{}%
\providecommand \@@endlink[0]{}%
\providecommand \url  [0]{\begingroup\@sanitize@url \@url }%
\providecommand \@url [1]{\endgroup\@href {#1}{\urlprefix }}%
\providecommand \urlprefix  [0]{URL }%
\providecommand \Eprint [0]{\href }%
\providecommand \doibase [0]{http://dx.doi.org/}%
\providecommand \selectlanguage [0]{\@gobble}%
\providecommand \bibinfo  [0]{\@secondoftwo}%
\providecommand \bibfield  [0]{\@secondoftwo}%
\providecommand \translation [1]{[#1]}%
\providecommand \BibitemOpen [0]{}%
\providecommand \bibitemStop [0]{}%
\providecommand \bibitemNoStop [0]{.\EOS\space}%
\providecommand \EOS [0]{\spacefactor3000\relax}%
\providecommand \BibitemShut  [1]{\csname bibitem#1\endcsname}%
\let\auto@bib@innerbib\@empty
\bibitem [{\citenamefont {Aigner}\ \emph {et~al.}(2016)\citenamefont {Aigner},
  \citenamefont {Wiesinger}, \citenamefont {Wiggers}, \citenamefont
  {Struztmann},\ and\ \citenamefont {Pereira}}]{Aigner}%
  \BibitemOpen
  \bibfield  {author} {\bibinfo {author} {\bibfnamefont {W.}~\bibnamefont
  {Aigner}}, \bibinfo {author} {\bibfnamefont {M.}~\bibnamefont {Wiesinger}},
  \bibinfo {author} {\bibfnamefont {H.}~\bibnamefont {Wiggers}}, \bibinfo
  {author} {\bibfnamefont {M.}~\bibnamefont {Struztmann}}, \ and\ \bibinfo
  {author} {\bibfnamefont {R.~N.}\ \bibnamefont {Pereira}},\ }\href@noop {}
  {\bibfield  {journal} {\bibinfo  {journal} {Physical Review Applied}\
  }\textbf {\bibinfo {volume} {5}},\ \bibinfo {pages} {054017} (\bibinfo {year}
  {2016})}\BibitemShut {NoStop}%
\bibitem [{\citenamefont {Li}\ \emph {et~al.}(2012)\citenamefont {Li},
  \citenamefont {Graf}, \citenamefont {Schladt}, \citenamefont {Jiang},\ and\
  \citenamefont {Parkin}}]{Li}%
  \BibitemOpen
  \bibfield  {author} {\bibinfo {author} {\bibfnamefont {M.}~\bibnamefont
  {Li}}, \bibinfo {author} {\bibfnamefont {T.}~\bibnamefont {Graf}}, \bibinfo
  {author} {\bibfnamefont {T.~D.}\ \bibnamefont {Schladt}}, \bibinfo {author}
  {\bibfnamefont {X.}~\bibnamefont {Jiang}}, \ and\ \bibinfo {author}
  {\bibfnamefont {S.~S.~P.}\ \bibnamefont {Parkin}},\ }\href@noop {} {\bibfield
   {journal} {\bibinfo  {journal} {Physical Review Letters}\ }\textbf {\bibinfo
  {volume} {109}},\ \bibinfo {pages} {196803} (\bibinfo {year}
  {2012})}\BibitemShut {NoStop}%
\bibitem [{\citenamefont {Sokolowska}\ \emph {et~al.}(2013)\citenamefont
  {Sokolowska}, \citenamefont {Dziob}, \citenamefont {Gorska}, \citenamefont
  {Kieltyka},\ and\ \citenamefont {Moscicki}}]{Sokolowska}%
  \BibitemOpen
  \bibfield  {author} {\bibinfo {author} {\bibfnamefont {D.}~\bibnamefont
  {Sokolowska}}, \bibinfo {author} {\bibfnamefont {D.}~\bibnamefont {Dziob}},
  \bibinfo {author} {\bibfnamefont {U.}~\bibnamefont {Gorska}}, \bibinfo
  {author} {\bibfnamefont {B.}~\bibnamefont {Kieltyka}}, \ and\ \bibinfo
  {author} {\bibfnamefont {J.~K.}\ \bibnamefont {Moscicki}},\ }\href@noop {}
  {\bibfield  {journal} {\bibinfo  {journal} {Physical Review E}\ }\textbf
  {\bibinfo {volume} {87}},\ \bibinfo {pages} {062404} (\bibinfo {year}
  {2013})}\BibitemShut {NoStop}%
\bibitem [{\citenamefont {Sherafati}\ \emph {et~al.}(2016)\citenamefont
  {Sherafati}, \citenamefont {Baldini}, \citenamefont {Malavasi},\ and\
  \citenamefont {Satpathy}}]{Sherafati}%
  \BibitemOpen
  \bibfield  {author} {\bibinfo {author} {\bibfnamefont {M.}~\bibnamefont
  {Sherafati}}, \bibinfo {author} {\bibfnamefont {M.}~\bibnamefont {Baldini}},
  \bibinfo {author} {\bibfnamefont {L.}~\bibnamefont {Malavasi}}, \ and\
  \bibinfo {author} {\bibfnamefont {S.}~\bibnamefont {Satpathy}},\ }\href@noop
  {} {\bibfield  {journal} {\bibinfo  {journal} {Physical Review B}\ }\textbf
  {\bibinfo {volume} {93}},\ \bibinfo {pages} {024107} (\bibinfo {year}
  {2016})}\BibitemShut {NoStop}%
\bibitem [{\citenamefont {Cowley}\ \emph {et~al.}(1980)\citenamefont {Cowley},
  \citenamefont {Birgeneau}, \citenamefont {Shirane}, \citenamefont
  {Guggenheim},\ and\ \citenamefont {Ikeda}}]{Cowley}%
  \BibitemOpen
  \bibfield  {author} {\bibinfo {author} {\bibfnamefont {R.~A.}\ \bibnamefont
  {Cowley}}, \bibinfo {author} {\bibfnamefont {R.~J.}\ \bibnamefont
  {Birgeneau}}, \bibinfo {author} {\bibfnamefont {G.}~\bibnamefont {Shirane}},
  \bibinfo {author} {\bibfnamefont {H.~J.}\ \bibnamefont {Guggenheim}}, \ and\
  \bibinfo {author} {\bibfnamefont {H.}~\bibnamefont {Ikeda}},\ }\href@noop {}
  {\bibfield  {journal} {\bibinfo  {journal} {Physical Review B}\ }\textbf
  {\bibinfo {volume} {21}},\ \bibinfo {pages} {4038} (\bibinfo {year}
  {1980})}\BibitemShut {NoStop}%
\bibitem [{\citenamefont {Birgeneau}\ \emph {et~al.}(1980)\citenamefont
  {Birgeneau}, \citenamefont {Cowley}, \citenamefont {Shirane}, \citenamefont
  {Tarvin},\ and\ \citenamefont {Guggenheim}}]{Birgeneau}%
  \BibitemOpen
  \bibfield  {author} {\bibinfo {author} {\bibfnamefont {R.~J.}\ \bibnamefont
  {Birgeneau}}, \bibinfo {author} {\bibfnamefont {R.~A.}\ \bibnamefont
  {Cowley}}, \bibinfo {author} {\bibfnamefont {G.}~\bibnamefont {Shirane}},
  \bibinfo {author} {\bibfnamefont {J.~A.}\ \bibnamefont {Tarvin}}, \ and\
  \bibinfo {author} {\bibfnamefont {H.~J.}\ \bibnamefont {Guggenheim}},\
  }\href@noop {} {\bibfield  {journal} {\bibinfo  {journal} {Physical Review
  B}\ }\textbf {\bibinfo {volume} {21}},\ \bibinfo {pages} {317} (\bibinfo
  {year} {1980})}\BibitemShut {NoStop}%
\bibitem [{\citenamefont {Birgeneau}\ \emph {et~al.}(1984)\citenamefont
  {Birgeneau}, \citenamefont {Cowley}, \citenamefont {Shirane},\ and\
  \citenamefont {Yoshizawa}}]{Birgeneau2}%
  \BibitemOpen
  \bibfield  {author} {\bibinfo {author} {\bibfnamefont {R.~J.}\ \bibnamefont
  {Birgeneau}}, \bibinfo {author} {\bibfnamefont {R.~A.}\ \bibnamefont
  {Cowley}}, \bibinfo {author} {\bibfnamefont {G.}~\bibnamefont {Shirane}}, \
  and\ \bibinfo {author} {\bibfnamefont {H.}~\bibnamefont {Yoshizawa}},\
  }\href@noop {} {\bibfield  {journal} {\bibinfo  {journal} {Journal of
  Statistical Physics}\ }\textbf {\bibinfo {volume} {34}},\ \bibinfo {pages}
  {817} (\bibinfo {year} {1984})}\BibitemShut {NoStop}%
\bibitem [{\citenamefont {Coniglio}(1981)}]{Coniglio}%
  \BibitemOpen
  \bibfield  {author} {\bibinfo {author} {\bibfnamefont {A.}~\bibnamefont
  {Coniglio}},\ }\href@noop {} {\bibfield  {journal} {\bibinfo  {journal}
  {Physical Review Letters}\ }\textbf {\bibinfo {volume} {46}},\ \bibinfo
  {pages} {250} (\bibinfo {year} {1981})}\BibitemShut {NoStop}%
\bibitem [{\citenamefont {Chinta}\ and\ \citenamefont
  {Headrik}(2014)}]{Chinta}%
  \BibitemOpen
  \bibfield  {author} {\bibinfo {author} {\bibfnamefont {P.~V.}\ \bibnamefont
  {Chinta}}\ and\ \bibinfo {author} {\bibfnamefont {R.~L.}\ \bibnamefont
  {Headrik}},\ }\href@noop {} {\bibfield  {journal} {\bibinfo  {journal}
  {Physical Review Letters}\ }\textbf {\bibinfo {volume} {112}},\ \bibinfo
  {pages} {075503} (\bibinfo {year} {2014})}\BibitemShut {NoStop}%
\bibitem [{\citenamefont {Sattar}\ \emph {et~al.}(2013)\citenamefont {Sattar},
  \citenamefont {Fostner},\ and\ \citenamefont {Brown}}]{Sattar}%
  \BibitemOpen
  \bibfield  {author} {\bibinfo {author} {\bibfnamefont {A.}~\bibnamefont
  {Sattar}}, \bibinfo {author} {\bibfnamefont {S.}~\bibnamefont {Fostner}}, \
  and\ \bibinfo {author} {\bibfnamefont {S.~A.}\ \bibnamefont {Brown}},\
  }\href@noop {} {\bibfield  {journal} {\bibinfo  {journal} {Physical Review
  Letters}\ }\textbf {\bibinfo {volume} {111}},\ \bibinfo {pages} {136808}
  (\bibinfo {year} {2013})}\BibitemShut {NoStop}%
\bibitem [{\citenamefont {Elmers}\ \emph {et~al.}(1994)\citenamefont {Elmers},
  \citenamefont {Hauschild}, \citenamefont {H{\"{o}}che}, \citenamefont
  {Gradmann}, \citenamefont {Bethge}, \citenamefont {Heuer},\ and\
  \citenamefont {K{\"{o}}hler}}]{Elmers}%
  \BibitemOpen
  \bibfield  {author} {\bibinfo {author} {\bibfnamefont {H.~J.}\ \bibnamefont
  {Elmers}}, \bibinfo {author} {\bibfnamefont {J.}~\bibnamefont {Hauschild}},
  \bibinfo {author} {\bibfnamefont {H.}~\bibnamefont {H{\"{o}}che}}, \bibinfo
  {author} {\bibfnamefont {U.}~\bibnamefont {Gradmann}}, \bibinfo {author}
  {\bibfnamefont {H.}~\bibnamefont {Bethge}}, \bibinfo {author} {\bibfnamefont
  {D.}~\bibnamefont {Heuer}}, \ and\ \bibinfo {author} {\bibfnamefont
  {U.}~\bibnamefont {K{\"{o}}hler}},\ }\href@noop {} {\bibfield  {journal}
  {\bibinfo  {journal} {Physical Review Letters}\ }\textbf {\bibinfo {volume}
  {73}},\ \bibinfo {pages} {898} (\bibinfo {year} {1994})}\BibitemShut
  {NoStop}%
\bibitem [{\citenamefont {Stauffer}\ and\ \citenamefont
  {Aharony}(1994)}]{StaufferAharony}%
  \BibitemOpen
  \bibfield  {author} {\bibinfo {author} {\bibfnamefont {D.}~\bibnamefont
  {Stauffer}}\ and\ \bibinfo {author} {\bibfnamefont {A.}~\bibnamefont
  {Aharony}},\ }\href@noop {} {\emph {\bibinfo {title} {Introduction to
  Percolation Theory}}},\ \bibinfo {edition} {2nd}\ ed.\ (\bibinfo  {publisher}
  {Taylor and Francis},\ \bibinfo {year} {1994})\BibitemShut {NoStop}%
\bibitem [{\citenamefont {Bovensiepen}\ \emph {et~al.}(1999)\citenamefont
  {Bovensiepen}, \citenamefont {Poulopoulos}, \citenamefont {Platow},
  \citenamefont {Farle},\ and\ \citenamefont {Babershke}}]{Bovensiepen}%
  \BibitemOpen
  \bibfield  {author} {\bibinfo {author} {\bibfnamefont {U.}~\bibnamefont
  {Bovensiepen}}, \bibinfo {author} {\bibfnamefont {P.}~\bibnamefont
  {Poulopoulos}}, \bibinfo {author} {\bibfnamefont {W.}~\bibnamefont {Platow}},
  \bibinfo {author} {\bibfnamefont {M.}~\bibnamefont {Farle}}, \ and\ \bibinfo
  {author} {\bibfnamefont {K.}~\bibnamefont {Babershke}},\ }\href@noop {}
  {\bibfield  {journal} {\bibinfo  {journal} {J. Magn. Magn. Mat.}\ }\textbf
  {\bibinfo {volume} {192}},\ \bibinfo {pages} {L386} (\bibinfo {year}
  {1999})}\BibitemShut {NoStop}%
\bibitem [{\citenamefont {K{\"{u}}pper}\ \emph {et~al.}(2007)\citenamefont
  {K{\"{u}}pper}, \citenamefont {Easton},\ and\ \citenamefont
  {Bland}}]{Kupper}%
  \BibitemOpen
  \bibfield  {author} {\bibinfo {author} {\bibfnamefont {D.}~\bibnamefont
  {K{\"{u}}pper}}, \bibinfo {author} {\bibfnamefont {S.}~\bibnamefont
  {Easton}}, \ and\ \bibinfo {author} {\bibfnamefont {J.~A.~C.}\ \bibnamefont
  {Bland}},\ }\href@noop {} {\bibfield  {journal} {\bibinfo  {journal} {J.
  Appl. Phys.}\ }\textbf {\bibinfo {volume} {102}},\ \bibinfo {pages} {083902}
  (\bibinfo {year} {2007})}\BibitemShut {NoStop}%
\bibitem [{\citenamefont {Hope}\ \emph {et~al.}(1999)\citenamefont {Hope},
  \citenamefont {Tselepi}, \citenamefont {Gu}, \citenamefont {Parker},\ and\
  \citenamefont {Bland}}]{Hope}%
  \BibitemOpen
  \bibfield  {author} {\bibinfo {author} {\bibfnamefont {S.}~\bibnamefont
  {Hope}}, \bibinfo {author} {\bibfnamefont {M.}~\bibnamefont {Tselepi}},
  \bibinfo {author} {\bibfnamefont {E.}~\bibnamefont {Gu}}, \bibinfo {author}
  {\bibfnamefont {T.~M.}\ \bibnamefont {Parker}}, \ and\ \bibinfo {author}
  {\bibfnamefont {J.~A.~C.}\ \bibnamefont {Bland}},\ }\href@noop {} {\bibfield
  {journal} {\bibinfo  {journal} {J. Appl. Phys.}\ }\textbf {\bibinfo {volume}
  {85}},\ \bibinfo {pages} {6094} (\bibinfo {year} {1999})}\BibitemShut
  {NoStop}%
\bibitem [{\citenamefont {Choi}\ \emph {et~al.}(1999)\citenamefont {Choi},
  \citenamefont {Kim}, \citenamefont {Schuller}, \citenamefont {Park},\ and\
  \citenamefont {Whang}}]{Choi}%
  \BibitemOpen
  \bibfield  {author} {\bibinfo {author} {\bibfnamefont {J.~M.}\ \bibnamefont
  {Choi}}, \bibinfo {author} {\bibfnamefont {S.}~\bibnamefont {Kim}}, \bibinfo
  {author} {\bibfnamefont {I.~K.}\ \bibnamefont {Schuller}}, \bibinfo {author}
  {\bibfnamefont {S.~M.}\ \bibnamefont {Park}}, \ and\ \bibinfo {author}
  {\bibfnamefont {C.~N.}\ \bibnamefont {Whang}},\ }\href@noop {} {\bibfield
  {journal} {\bibinfo  {journal} {J. Magn. Magn. Mat.}\ }\textbf {\bibinfo
  {volume} {191}},\ \bibinfo {pages} {54} (\bibinfo {year} {1999})}\BibitemShut
  {NoStop}%
\bibitem [{\citenamefont {Hauschild}\ \emph {et~al.}(1998)\citenamefont
  {Hauschild}, \citenamefont {Elmers},\ and\ \citenamefont
  {Gradmann}}]{Hauschild}%
  \BibitemOpen
  \bibfield  {author} {\bibinfo {author} {\bibfnamefont {J.}~\bibnamefont
  {Hauschild}}, \bibinfo {author} {\bibfnamefont {H.~J.}\ \bibnamefont
  {Elmers}}, \ and\ \bibinfo {author} {\bibfnamefont {U.}~\bibnamefont
  {Gradmann}},\ }\href@noop {} {\bibfield  {journal} {\bibinfo  {journal}
  {Physical Review B}\ }\textbf {\bibinfo {volume} {57}},\ \bibinfo {pages}
  {R677} (\bibinfo {year} {1998})}\BibitemShut {NoStop}%
\bibitem [{\citenamefont {Pratzer}\ \emph {et~al.}(2001)\citenamefont
  {Pratzer}, \citenamefont {Elmers}, \citenamefont {Bode}, \citenamefont
  {Pietzsch}, \citenamefont {Kubetzka},\ and\ \citenamefont
  {Wiesendanger}}]{Pratzer}%
  \BibitemOpen
  \bibfield  {author} {\bibinfo {author} {\bibfnamefont {M.}~\bibnamefont
  {Pratzer}}, \bibinfo {author} {\bibfnamefont {H.~J.}\ \bibnamefont {Elmers}},
  \bibinfo {author} {\bibfnamefont {M.}~\bibnamefont {Bode}}, \bibinfo {author}
  {\bibfnamefont {O.}~\bibnamefont {Pietzsch}}, \bibinfo {author}
  {\bibfnamefont {A.}~\bibnamefont {Kubetzka}}, \ and\ \bibinfo {author}
  {\bibfnamefont {R.}~\bibnamefont {Wiesendanger}},\ }\href@noop {} {\bibfield
  {journal} {\bibinfo  {journal} {Physical Review Letters}\ }\textbf {\bibinfo
  {volume} {87}},\ \bibinfo {pages} {127201} (\bibinfo {year}
  {2001})}\BibitemShut {NoStop}%
\bibitem [{\citenamefont {Elmers}\ \emph
  {et~al.}(1995{\natexlab{a}})\citenamefont {Elmers}, \citenamefont
  {Hauschild},\ and\ \citenamefont {Gradmann}}]{Elmers3}%
  \BibitemOpen
  \bibfield  {author} {\bibinfo {author} {\bibfnamefont {H.~J.}\ \bibnamefont
  {Elmers}}, \bibinfo {author} {\bibfnamefont {J.}~\bibnamefont {Hauschild}}, \
  and\ \bibinfo {author} {\bibfnamefont {U.}~\bibnamefont {Gradmann}},\
  }\href@noop {} {\bibfield  {journal} {\bibinfo  {journal} {J. Magn. Magn.
  Mat.}\ }\textbf {\bibinfo {volume} {140-144}},\ \bibinfo {pages} {1559}
  (\bibinfo {year} {1995}{\natexlab{a}})}\BibitemShut {NoStop}%
\bibitem [{\citenamefont {Weber}\ \emph {et~al.}(1997)\citenamefont {Weber},
  \citenamefont {Wagner}, \citenamefont {Elmers}, \citenamefont {Hauschild},\
  and\ \citenamefont {Gradmann}}]{Weber}%
  \BibitemOpen
  \bibfield  {author} {\bibinfo {author} {\bibfnamefont {N.}~\bibnamefont
  {Weber}}, \bibinfo {author} {\bibfnamefont {K.}~\bibnamefont {Wagner}},
  \bibinfo {author} {\bibfnamefont {H.~J.}\ \bibnamefont {Elmers}}, \bibinfo
  {author} {\bibfnamefont {J.}~\bibnamefont {Hauschild}}, \ and\ \bibinfo
  {author} {\bibfnamefont {U.}~\bibnamefont {Gradmann}},\ }\href@noop {}
  {\bibfield  {journal} {\bibinfo  {journal} {Physical Review B}\ }\textbf
  {\bibinfo {volume} {55}},\ \bibinfo {pages} {14121} (\bibinfo {year}
  {1997})}\BibitemShut {NoStop}%
\bibitem [{\citenamefont {Elmers}\ \emph
  {et~al.}(1995{\natexlab{b}})\citenamefont {Elmers}, \citenamefont
  {Hauschild}, \citenamefont {Fritzsche}, \citenamefont {Liu}, \citenamefont
  {Gradmann},\ and\ \citenamefont {K{\"{o}}hler}}]{Elmers2}%
  \BibitemOpen
  \bibfield  {author} {\bibinfo {author} {\bibfnamefont {H.~J.}\ \bibnamefont
  {Elmers}}, \bibinfo {author} {\bibfnamefont {J.}~\bibnamefont {Hauschild}},
  \bibinfo {author} {\bibfnamefont {H.}~\bibnamefont {Fritzsche}}, \bibinfo
  {author} {\bibfnamefont {G.}~\bibnamefont {Liu}}, \bibinfo {author}
  {\bibfnamefont {U.}~\bibnamefont {Gradmann}}, \ and\ \bibinfo {author}
  {\bibfnamefont {U.}~\bibnamefont {K{\"{o}}hler}},\ }\href@noop {} {\bibfield
  {journal} {\bibinfo  {journal} {Physical Review Letters}\ }\textbf {\bibinfo
  {volume} {75}},\ \bibinfo {pages} {2031} (\bibinfo {year}
  {1995}{\natexlab{b}})}\BibitemShut {NoStop}%
\bibitem [{\citenamefont {Gradmann}\ and\ \citenamefont
  {Waller}(1982)}]{Gradmann}%
  \BibitemOpen
  \bibfield  {author} {\bibinfo {author} {\bibfnamefont {U.}~\bibnamefont
  {Gradmann}}\ and\ \bibinfo {author} {\bibfnamefont {G.}~\bibnamefont
  {Waller}},\ }\href@noop {} {\bibfield  {journal} {\bibinfo  {journal}
  {Surface Science}\ }\textbf {\bibinfo {volume} {116}},\ \bibinfo {pages}
  {539} (\bibinfo {year} {1982})}\BibitemShut {NoStop}%
\bibitem [{\citenamefont {Herring}\ and\ \citenamefont
  {Kittel}(1951)}]{Herring}%
  \BibitemOpen
  \bibfield  {author} {\bibinfo {author} {\bibfnamefont {C.}~\bibnamefont
  {Herring}}\ and\ \bibinfo {author} {\bibfnamefont {C.}~\bibnamefont
  {Kittel}},\ }\href@noop {} {\bibfield  {journal} {\bibinfo  {journal}
  {Physical Review B}\ }\textbf {\bibinfo {volume} {81}},\ \bibinfo {pages}
  {869} (\bibinfo {year} {1951})}\BibitemShut {NoStop}%
\bibitem [{\citenamefont {Bander}\ and\ \citenamefont {Mills}(1988)}]{Bander}%
  \BibitemOpen
  \bibfield  {author} {\bibinfo {author} {\bibfnamefont {M.}~\bibnamefont
  {Bander}}\ and\ \bibinfo {author} {\bibfnamefont {D.~L.}\ \bibnamefont
  {Mills}},\ }\href@noop {} {\bibfield  {journal} {\bibinfo  {journal}
  {Physical Review B}\ }\textbf {\bibinfo {volume} {38}},\ \bibinfo {pages}
  {12015} (\bibinfo {year} {1988})}\BibitemShut {NoStop}%
\bibitem [{\citenamefont {Kramers}\ and\ \citenamefont
  {Wannier}(1941)}]{Kramers}%
  \BibitemOpen
  \bibfield  {author} {\bibinfo {author} {\bibfnamefont {H.~A.}\ \bibnamefont
  {Kramers}}\ and\ \bibinfo {author} {\bibfnamefont {G.~H.}\ \bibnamefont
  {Wannier}},\ }\href@noop {} {\bibfield  {journal} {\bibinfo  {journal}
  {Physical Review}\ }\textbf {\bibinfo {volume} {60}},\ \bibinfo {pages} {252}
  (\bibinfo {year} {1941})}\BibitemShut {NoStop}%
\bibitem [{\citenamefont {Serena}\ \emph {et~al.}(1993)\citenamefont {Serena},
  \citenamefont {Garc{\'{i}}a},\ and\ \citenamefont {Levanyuk}}]{Serena}%
  \BibitemOpen
  \bibfield  {author} {\bibinfo {author} {\bibfnamefont {P.~A.}\ \bibnamefont
  {Serena}}, \bibinfo {author} {\bibfnamefont {N.}~\bibnamefont
  {Garc{\'{i}}a}}, \ and\ \bibinfo {author} {\bibfnamefont {A.}~\bibnamefont
  {Levanyuk}},\ }\href@noop {} {\bibfield  {journal} {\bibinfo  {journal}
  {Physical Review B}\ }\textbf {\bibinfo {volume} {47}},\ \bibinfo {pages}
  {5027} (\bibinfo {year} {1993})}\BibitemShut {NoStop}%
\bibitem [{\citenamefont {Back}\ \emph {et~al.}(1995)\citenamefont {Back},
  \citenamefont {W{\"{u}}rsch}, \citenamefont {Vaterlaus}, \citenamefont
  {Ramsperger}, \citenamefont {Maier},\ and\ \citenamefont {Pescia}}]{Back}%
  \BibitemOpen
  \bibfield  {author} {\bibinfo {author} {\bibfnamefont {C.~H.}\ \bibnamefont
  {Back}}, \bibinfo {author} {\bibfnamefont {C.}~\bibnamefont {W{\"{u}}rsch}},
  \bibinfo {author} {\bibfnamefont {A.}~\bibnamefont {Vaterlaus}}, \bibinfo
  {author} {\bibfnamefont {U.}~\bibnamefont {Ramsperger}}, \bibinfo {author}
  {\bibfnamefont {U.}~\bibnamefont {Maier}}, \ and\ \bibinfo {author}
  {\bibfnamefont {D.}~\bibnamefont {Pescia}},\ }\href@noop {} {\bibfield
  {journal} {\bibinfo  {journal} {Nature}\ }\textbf {\bibinfo {volume} {378}},\
  \bibinfo {pages} {597} (\bibinfo {year} {1995})}\BibitemShut {NoStop}%
\bibitem [{\citenamefont {Dunlavy}\ and\ \citenamefont
  {Venus}(2004)}]{Dunlavy}%
  \BibitemOpen
  \bibfield  {author} {\bibinfo {author} {\bibfnamefont {M.~J.}\ \bibnamefont
  {Dunlavy}}\ and\ \bibinfo {author} {\bibfnamefont {D.}~\bibnamefont
  {Venus}},\ }\href@noop {} {\bibfield  {journal} {\bibinfo  {journal}
  {Physical Review B}\ }\textbf {\bibinfo {volume} {69}},\ \bibinfo {pages}
  {094411} (\bibinfo {year} {2004})}\BibitemShut {NoStop}%
\bibitem [{\citenamefont {Venus}(1995)}]{Venus}%
  \BibitemOpen
  \bibfield  {author} {\bibinfo {author} {\bibfnamefont {D.}~\bibnamefont
  {Venus}},\ }\href@noop {} {\bibfield  {journal} {\bibinfo  {journal} {Rev.
  Sci. Instum.}\ }\textbf {\bibinfo {volume} {66}},\ \bibinfo {pages} {3280}
  (\bibinfo {year} {1995})}\BibitemShut {NoStop}%
\bibitem [{\citenamefont {Jones}\ \emph {et~al.}(1993)\citenamefont {Jones},
  \citenamefont {Sawler},\ and\ \citenamefont {Venus}}]{Jones}%
  \BibitemOpen
  \bibfield  {author} {\bibinfo {author} {\bibfnamefont {T.~L.}\ \bibnamefont
  {Jones}}, \bibinfo {author} {\bibfnamefont {J.}~\bibnamefont {Sawler}}, \
  and\ \bibinfo {author} {\bibfnamefont {D.}~\bibnamefont {Venus}},\
  }\href@noop {} {\bibfield  {journal} {\bibinfo  {journal} {Rev. Sci.
  Instrum.}\ }\textbf {\bibinfo {volume} {64}},\ \bibinfo {pages} {2008}
  (\bibinfo {year} {1993})}\BibitemShut {NoStop}%
\bibitem [{\citenamefont {Fritsch}\ \emph {et~al.}(2011)\citenamefont
  {Fritsch}, \citenamefont {D'Ortenzio},\ and\ \citenamefont
  {Venus}}]{Fritsch}%
  \BibitemOpen
  \bibfield  {author} {\bibinfo {author} {\bibfnamefont {K.}~\bibnamefont
  {Fritsch}}, \bibinfo {author} {\bibfnamefont {R.}~\bibnamefont {D'Ortenzio}},
  \ and\ \bibinfo {author} {\bibfnamefont {D.}~\bibnamefont {Venus}},\
  }\href@noop {} {\bibfield  {journal} {\bibinfo  {journal} {Physical Review
  B}\ }\textbf {\bibinfo {volume} {83}},\ \bibinfo {pages} {075421} (\bibinfo
  {year} {2011})}\BibitemShut {NoStop}%
\bibitem [{\citenamefont {He}\ \emph {et~al.}(2016)\citenamefont {He},
  \citenamefont {Winch}, \citenamefont {Belanger}, \citenamefont {Nguyen},\
  and\ \citenamefont {Venus}}]{He}%
  \BibitemOpen
  \bibfield  {author} {\bibinfo {author} {\bibfnamefont {G.}~\bibnamefont
  {He}}, \bibinfo {author} {\bibfnamefont {H.}~\bibnamefont {Winch}}, \bibinfo
  {author} {\bibfnamefont {R.}~\bibnamefont {Belanger}}, \bibinfo {author}
  {\bibfnamefont {P.}~\bibnamefont {Nguyen}}, \ and\ \bibinfo {author}
  {\bibfnamefont {D.}~\bibnamefont {Venus}},\ }\href@noop {} {} (\bibinfo
  {year} {2016}),\ \bibinfo {note} {arxiv:1610.06446 (submitted)}\BibitemShut
  {NoStop}%
\bibitem [{\citenamefont {Arnold}\ \emph {et~al.}(1997)\citenamefont {Arnold},
  \citenamefont {Dunlavy},\ and\ \citenamefont {Venus}}]{Arnold}%
  \BibitemOpen
  \bibfield  {author} {\bibinfo {author} {\bibfnamefont {C.~S.}\ \bibnamefont
  {Arnold}}, \bibinfo {author} {\bibfnamefont {M.}~\bibnamefont {Dunlavy}}, \
  and\ \bibinfo {author} {\bibfnamefont {D.}~\bibnamefont {Venus}},\
  }\href@noop {} {\bibfield  {journal} {\bibinfo  {journal} {Rev. Sci.
  Instrum.}\ }\textbf {\bibinfo {volume} {68}},\ \bibinfo {pages} {4212}
  (\bibinfo {year} {1997})}\BibitemShut {NoStop}%
\bibitem [{\citenamefont {D{\"{u}}rkop}\ \emph {et~al.}(1997)\citenamefont
  {D{\"{u}}rkop}, \citenamefont {Elmers},\ and\ \citenamefont
  {Gradmann}}]{Durkop}%
  \BibitemOpen
  \bibfield  {author} {\bibinfo {author} {\bibfnamefont {T.}~\bibnamefont
  {D{\"{u}}rkop}}, \bibinfo {author} {\bibfnamefont {H.~J.}\ \bibnamefont
  {Elmers}}, \ and\ \bibinfo {author} {\bibfnamefont {U.}~\bibnamefont
  {Gradmann}},\ }\href@noop {} {\bibfield  {journal} {\bibinfo  {journal} {J.
  Mag. Mag. Mat.}\ }\textbf {\bibinfo {volume} {172}},\ \bibinfo {pages} {L1}
  (\bibinfo {year} {1997})}\BibitemShut {NoStop}%
\bibitem [{\citenamefont {Elmers}\ \emph {et~al.}(1999)\citenamefont {Elmers},
  \citenamefont {Hauschild},\ and\ \citenamefont {Gradmann}}]{Elmers5}%
  \BibitemOpen
  \bibfield  {author} {\bibinfo {author} {\bibfnamefont {H.~J.}\ \bibnamefont
  {Elmers}}, \bibinfo {author} {\bibfnamefont {J.}~\bibnamefont {Hauschild}}, \
  and\ \bibinfo {author} {\bibfnamefont {U.}~\bibnamefont {Gradmann}},\
  }\href@noop {} {\bibfield  {journal} {\bibinfo  {journal} {Physical Review
  B}\ }\textbf {\bibinfo {volume} {59}},\ \bibinfo {pages} {3688} (\bibinfo
  {year} {1999})}\BibitemShut {NoStop}%
\bibitem [{\citenamefont {Elmers}\ and\ \citenamefont
  {Gradmann}(1990)}]{Elmers4}%
  \BibitemOpen
  \bibfield  {author} {\bibinfo {author} {\bibfnamefont {H.~J.}\ \bibnamefont
  {Elmers}}\ and\ \bibinfo {author} {\bibfnamefont {U.}~\bibnamefont
  {Gradmann}},\ }\href@noop {} {\bibfield  {journal} {\bibinfo  {journal}
  {Appl. Phys. A}\ }\textbf {\bibinfo {volume} {51}},\ \bibinfo {pages} {255}
  (\bibinfo {year} {1990})}\BibitemShut {NoStop}%
\end{thebibliography}%

\end{document}